# Network Intrusion Detection Using FP Tree Rules

**P Srinivasulu, J Ranga Rao**

Department of Computer Science and Engineering, V R Siddhartha Engineering College, Vijayawada

**I Ramesh Babu**

Department of Computer Science and Engineering, Acharya  Nagarjuna University,  Guntur

-----------------------------------------------ABSTRACT-----------------------------------------------
In the faceless world of the Internet, online fraud is one of the greatest reasons of loss for web merchants. Advanced solutions are needed to protect e-businesses from the constant problems of fraud. Many popular fraud detection algorithms require supervised training, which needs human intervention to prepare training cases. Since it is quite often for an online transaction database to have Terabyte-level storage, human investigation to identify fraudulent transactions is very costly. This paper describes the automatic design of user profiling method for the purpose of fraud detection. We use *a FP (Frequent Pattern) Tree* rule-learning algorithm to adaptively profile legitimate customer behavior in a transaction database. Then the incoming transactions are compared against the user profile to uncover the *anomalies*. The anomaly outputs are used as input to an accumulation system for combining evidence to generate high-confidence fraud alert value. Favorable experimental results are presented.

## 1. INTRODUCTION

In anomaly detection, the goal is to find objects that are different from most other objects. Often anomalous, objects are known as *outliers*, since, on scatter plot of the data, they lie far away from other data points. Anomaly detection is also known as *deviation detection*, because anomalous objects have attribute values that deviate significantly from the expected or typical attribute values, or as *exception min*ing, because anomalies are exceptional in some sense.

There are a variety of anomaly detection approaches from several areas, including statistics, machine learning, and data mining. All try to capture the idea that an anomalous data object is unusual or in some way inconsistent with other objects. Our method for detecting fraud is to check for suspicious changes in user behavior. This paper describes the automatic design of user profiling methods for the purpose of fraud detection, using Frequent Pattern tree data mining techniques. Specifically, we use a rule-learning [1, 2, 3] program to uncover indicators of fraudulent behavior from a large database of customer transactions. Then the indicators are used to create a set of monitors, which profile legitimate customer behavior and indicate anomalies. Finally, the outputs of the monitors are used as features in a system that learns to combine evidence to generate high-confidence alarms. The system has been applied to the problem of detecting cellular cloning fraud based on a database of call records.

Experiments indicate that this automatic approach performs better than hand-crafted methods for detecting fraud. Furthermore, this approach can adapt to the changing conditions typical of fraud detection environments.

A novel fraud detection framework is proposed in this paper. Individual user's behavior pattern is dynamically profile from the transactions by using a set of association rules. The association rule [1] is first introduced by Agarwal. The incoming transactions for that user are then compared against the profile in order to discover the anomalies, based on which the corresponding warnings are outputted. Our algorithm is evaluated on both synthetic data and real data. An experimental results shows that our algorithm can accurately differentiate the anomaly behavior from profile user behavior.

## 2. THE BASIC IDEA

In this section, we will describe the basic idea of our fraud detection algorithm. Before doing so, we will first give some definitions.

*Definition 2.1.* A Set of attribute-value pairs or items $\sum$ = $\{a_i,(v_j)\}$, where i € {1, 2, ….., n }, n is the number of all possible attributes in the database we want to keep record, j € { 1, 2, .., m(i)}, m(i) is the possible values of attribute $a_i$, m(i) is depended on the granularity specified along the attribute $a_i$. We also use *Ii* to represent an attribute-value pair or item *ai (vj)* for simplicity.  An  example  of attribute-value pair is Price ("*$1-$10*"), where Price is an attribute, "*$1-$10*" is a value of this attribute. The possible values are depended on the granularity or interval of the attribute Price. The interval is $10 in this example. Another example of an attribute-value pair is Time ("*Evening*"), where Time is an attribute, "*Evening*" is a possible value. By different granularity the attribute-value pair could be Time("*9pm*"). The different granularity could cause large differences on the performance of behavior profiling.

*Definition 2.2.* The *transactions* are records of the form $T(t)$ where $t$ is a value of the time variable $D$. Each transaction consists of a set of certain attribute-value pairs from $\Sigma$ recorded in a period of $t$. A *transaction database* contains all the transactions.

The most recent transactions for an individual user in a transaction database are analyzed in order to profile the current behavior or habit for that customer. The word 'recent' is spelled by a slide window, which could be a time window or a transaction count window. For example, recent transactions could be all the transactions in the past two months, or the recent 500 transactions. The customer's profile is utilized to monitor a new transaction of this user to indicate how unusual the new transaction is. At the same time, the customer's old profile is automatically updated by accumulating the occurrences of the new attributes, which represents the user's new behavior. It is quite reasonable to assume that a normal user should have a behavior pattern which indicates his or hers consuming interests or habits, since a totally randomly consuming behavior is very uncommon.

We use a set of *association rules* to profile a user's recent behaviour.

*Definition 2.3. Association rules* [4] are the implication of the form $X \rightarrow Y$, where

1.           $X \subseteq \Sigma,\ Y \subseteq \Sigma.$
2.           $X \cap Y = \varnothing$
3.           $\exists I_i \in X, 1 \leq i \leq n$
4.           $\exists I_j \in Y, 1 \leq j \leq n$

The association rule $X \rightarrow Y$ is interpreted as data set that satisfies the conditions in $X$ are also likely to satisfy the conditions in $Y$. An example of an association rule is:day("*Saturday*")^time("*8pm10pm*")→play("*Xbox contest*") [*support*=15%, *confidence*=77%].

The rule indicates that for all transactions of a customer recorded in a time window, 15% (*support*) transactions are playing "*Xbox contest*" in "*Saturday*", "*8pm-10pm*". There is a 77% probability (*confidence*) that if a transaction happens in "*Saturday*", "*8pm-10pm*" it would be an Xbox contest.

Two important measures for association rules, *support* and *confidence*, are defined as follows.

*Definition 2.4.* The *support*, $s$, of an association rule is the ratio (in percent) of the transactions containing $X \cup Y$ to the total number of transactions analyzed, $|R(t)|$. If the *support* of an association rule is 15% then it means that 15% of the analyzed transactions contain $X \cup Y$. *Support* is the statistic significance of an association rule. The association rules have the *supports* less than 5% would be considered not very important to profile a user's behavior. While a high support is often desirable for association rules.

*Definition 2.5.* For a given number of transactions, *confidence*, $c$, is the ratio (in percent) of the number of transactions that contain $X \cup Y$ to the number of transactions that contain $X$. Thus if we say an association rule has a confidence of 77%, it means that 77% of the transactions containing $X$ also contain $Y$. The *confidence* of a rule indicates the degree of correlation in the dataset between $X$ and $Y$. It is used as a measure of a rule's strength. Often a large *confidence* is required for association rules. Considered not very important to profile a user's behavior. While a high *support* is often desirable for association rules.

An FP-tree [11] (frequent pattern tree) structure and FP-tree growth algorithm, proposed by Han are utilized to uncover these hidden association rules from the recent transactions for this user. We improve the FP-tree growth algorithm to allow it to mine both *intra-transaction associations* and the *inter-transaction association*.

Any new transaction of a user is compared against his or hers FP-tree to indicate the anomaly, which means how unusual the transaction is. A novel FP-tree based similarity measure algorithm is utilized to calculate the anomaly. We also use an accumulating algorithm to accumulate the low suspicious level warning to generate a high-confidence alarm. By comparing the alarm against a set of thresholds, corresponding fraud resolution could be performed.

## 3. NMT FRAUD DETECTION FRAMEWORK

We will describe the architecture of our experimental system, NMT (New Mexico Tech) fraud detection framework in this section. Our system consists of three major modules: Data engine, rule engine and rule monitor. They are shown inside a dashed rectangle in Figure 1. The objects surrounding the FDS (fraud detection system) make a typical online transaction system, which includes several web applications and web services to provide OLTP (OnLine Transaction Process), a database storing transaction data, and a database replication in order to provide minimum performance degradation on OLTP by backend data process or analysis.

FDS is a backend process, whose impact on the front end of the online system is minimized, since it only talks to the replication database. Data engine serves as an interface between the replication database and the FDS. It collects and pre-formats the recent transactions of all individual customers in the online system. The recent transactions depend on the detection sensitivity, which is specified by the customer through the web applications in OLTP. The rule engine module mines the recent transactions to generate a profile, an association rule set stored in an FP-tree, for each user. The FP-tree is updated adaptively after the new transactions of the user are added in replication database. The rule monitor module monitors the new transaction for every user. Any new transaction of a particular user is compared against the FP-tree for that user to indicate the anomaly. The anomaly is then mapped into a corresponding suspicious level, which is sent back to OLTP. The corresponding resolution is performed based on the suspicious level of a new transaction. In the following sub-sections of this paper, we will describe the implementation of rule engine and rule monitor in detail.

## 3.1 ADAPTIVE ASSOCIATION RULE MINING

The major responsibility of rule engine is to adaptively generate association rule sets to profile user behaviors. There are a large amount of techniques to mine association rules from a transaction

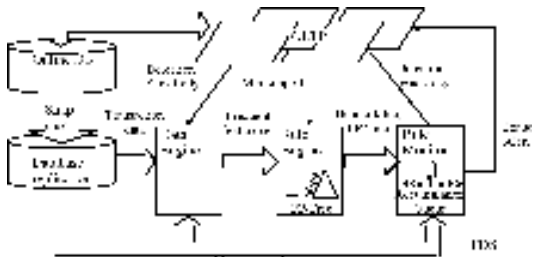

*Figure 1: Architecture of NMT fraud detection system.*

Database [6]. Agrawal introduced an Apriori-like mining method to mining association rules. An extension version of the Apriori algorithm [2] also by Agrawal is able to mine generalized, multilevel, or quantitative association rules [3]. Moe introduced association rule mining query languages. Ng introduced a constraint-based rule mining technique [16]. Cheung presented an incremental updating technique to discover the association rules in a large scale database. Agrawal proposed another technique to perform parallel and distributed mining. Brin introduced a dynamic itemset counting technique to reduce the number of database scans [4]. The most efficient association rule mining algorithm so far is the FP-tree growth algorithm [11] proposed by Han. It is able to mine the frequent patterns without candidate generation. Our fraud detection approach is largely based on this algorithm. The benefits of this method are highly condensed yet complete for frequent pattern mining, avoiding costly database scanning. More importantly, it allows us to adaptively generate user profile without preparing labeled data

| TID | Transaction Data | Frequent Item Sets |
|---|---|---|
| 1 | ET, ST, EV, 129.138, L50 | ST, 129.138, EV, ET |
| 2 | ET, ST, MR, 202.55, L10 | ST, ET, L10 |
| 3 | ET, SU, MR, 129.138, L50 | 129.138, ET |
| 4 | BK, ST, EV, 129.138, L10 | ST, 129.138, EV, L10 |
| 5 | CL, ST, EV, 129.138, L10 | ST, 129.138, EV, L10 |

*Table 1: An example of recent transactions for a customer. This table shows five transactions for a typical online transaction system. The right column lists the corresponding frequent itemsets. Minimum support value in this example is 60%. Minimum support is a user specified threshold for mining association rule from a database.*

Table 1 shows an example of five transactions. Each transaction is an itemset, which contains five items or attribute-value pairs.

*Definition 3.1.* A set of items or attribute-value pairs is referred to as an *itemset*. The occurrence *frequency* of an itemset is the number of transactions that contain the itemset. This is also known as *support count*. The frequency of itemset {ST} is 4 in this example.

*Definition 3.2.* If the occurrence frequency of an itemset is greater than or equal to the product of *min_sup* and the total number of transactions, then it is a *frequent itemsets*. The value of *min_sup* is called *minimum support* value. In the example shown in Table 1, the corresponding frequent itemset for transaction 2 is {ST, ET, L10}, since items MR, 202.55 are less than the minimum support value, 3 in this example (*min_sup* = 60%). Since the rarely occurred items would be filtered out when we generate the frequent itemsets, the rarely occurred behaviour such as fraudulent actions will be filtered out. That is the reason we could profile a user's behaviour without the preparation of the labeled data.

To profile the behaviour or a user, an FP-tree structure, introduced [13] by, is used to store compressed, crucial information about frequent pattern, from which association rules are generated. FP-tree is a combination of a general prefix tree and a linked list table. Figure 2 shows an example of creating a small FP-tree structure from the transactions shown in Table 1.

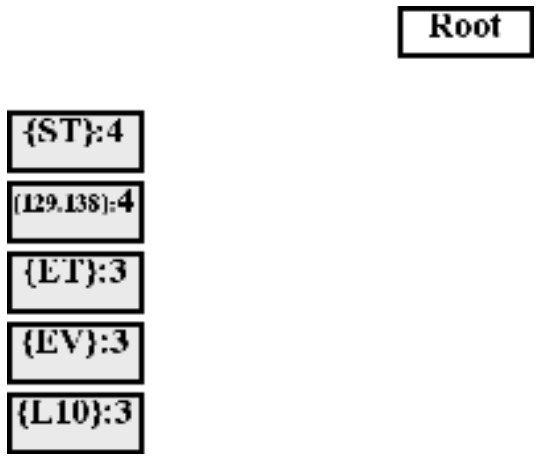

*(a) An empty FP-tree*

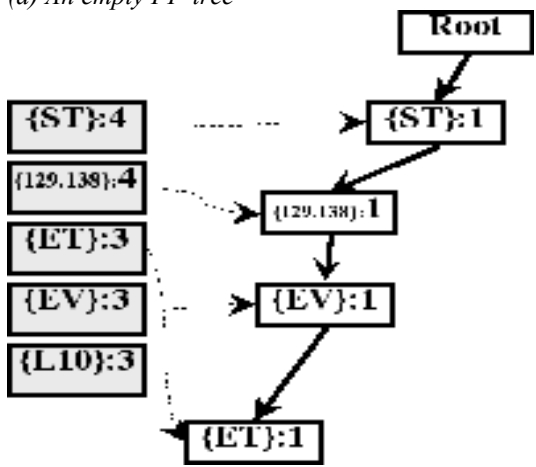

*(b) After inserting {ST, 129.138, EV, ET}*

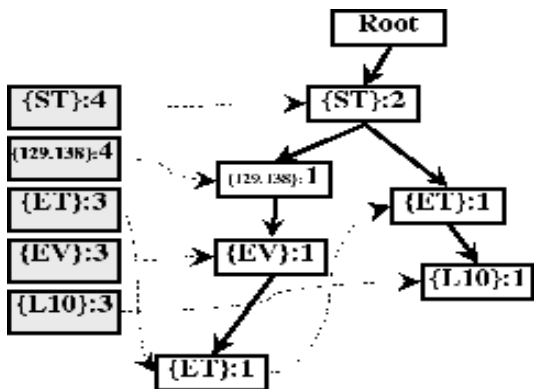

*(c) After inserting {ST, ET, L10}*

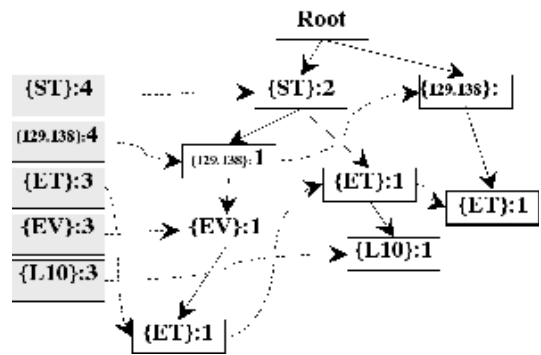

*(d) After inserting {129.138, ET}*

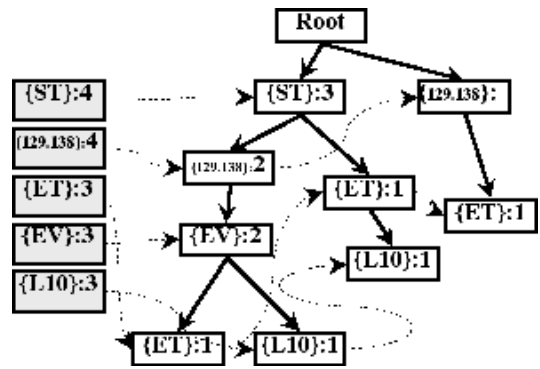

*(e) After inserting 1st {ST, 129.138, EV, L10}*

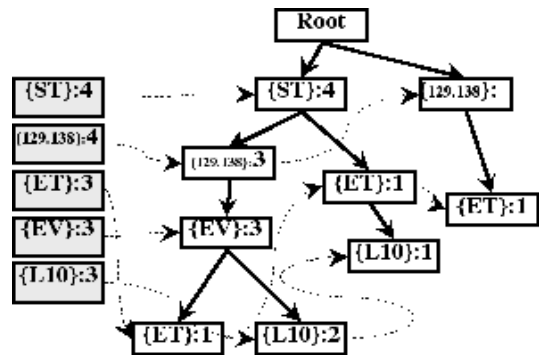

*(f) Insert 2nd {ST, 129.138, EV, L10}*

*Figure 2: An example of an FP-tree construction*

The FP-tree stores quantitative information about frequent pattern. The tree nodes are arranged in such a way that more frequently occurring nodes will have a better chance of sharing nodes than lesser ones. From every tree node to the root, the path is a frequent pattern, from which association rules could be generated.

FP-tree is constructed by accumulating the occurrences of the attributes in recent transactions of an individual customer. By mining an FP-tree, we can find all the conditional rules that correlate the presence of one set of features with that of another set of features. However, we are not interested in extracting these association rules. We use a pattern matching algorithm to compare a new transaction directly with the FP-tree to indicate the new transaction's anomaly. It will significantly improve the overall system performance. The most attractive feature of the FP-tree is that it could be updated by accumulating the occurrences of the attributes without any human involvement. It is an ideal data structure to profile and monitor a non-stationary data set.

To create an FP-tree, recent transactions should be scanned twice. The first scan of the transactions derives the set of frequent items. For example, the frequent itemset of the last transaction is {CL, ST, EV, 129.138, L10}. The frequent items are then sorted in the order of a descending support count. The resulting set or list is denoted *L*, which is {ST, 129.138, EV, L10} in this example *L* is utilized to construct a header table shown in the left side of Figure 2(a).

An FP-tree is then constructed from an empty root, shown in Figure 2(a). Branches of the FPtree are then inserted into the tree by scanning the transactions a second time. The items in each transaction are processed in *L* order, and a branch is created for each transaction. For example, the scan of the first transaction, "ET, ST, EV, 129.138, L50", which contains four items {ST, 129.138, EV, ET} in *L* order, shown in the first data row of Table 1, leads to the construction of the first branch of the tree with four tree nodes: {ST}:1,{129.138}:1,{EV}:1 and{ET}:1, where {ST}:1 is linked as a child of the root, {129.138}:1 is linked to {ST}:1…, shown in Figure 2(b). The number after the semicolon in the tree node shows the current occurrence of an item. To make tree traversal easy, an item header table is built so that each item pointer points to its occurrences in the tree via a chain of node links. The items in the header table are absolutely the same as the element in *L*.

The second transaction is processed in the same way as the first one. *T*2, "ET, ST, MR, 202.55, L10" contains {ST}, {ET} and {L10} in *L* order. These items would result in a branch where {ST}:1 is linked to the root and {ET}:1 is linked to {ST}:1, {L10}:1 is linked to {ET}:1. However this branch would share a common prefix, {ST}, with the existing path for *T*1. Therefore we should increase the count of the {ST} node by 1, and create two new tree nodes, {ET}:1 and {L10}:1, which is linked as a child tree of {ST}:2, shown in Figure 2(c). In general, when considering the branch to be added for a transaction, the count of each node along a common prefix is increased by 1, and nodes for the items following the prefix are created and linked accordingly. The repeated insertion for *T*3 – *T*5 are shown in Figures 2(d)–(f).

The size of an FP-tree is largely dependent on the minimum support, which decides the size of the frequent itemset for a transaction after filtering. The FP-tree is constructed from these itemsets. Therefore, the smaller minimum support value leads to a larger FP-tree, which profiles a user's behaviour more accurately, since it stores some associations having relatively small support counts.

However it requires more time to construct. The tradeoff between accurate profiling and time for tree construction is shown in Figure 3.

We can get all association rules under the condition of a particular frequency item by following its node link chain in header table. For example, to get all rules for {L10} we should firstly search the header table to find the node link chain of {L10}. The first node in the link chain is {L10}:2, whose prefix path is ({ST}:4, {129.138}:3, {EV}:3). The corresponding association is

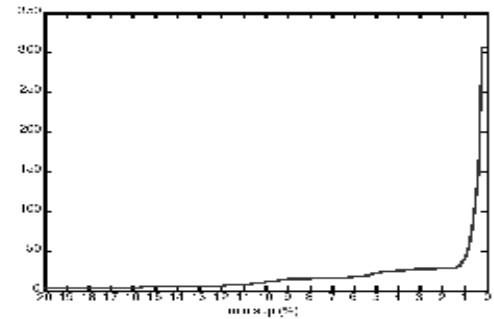

**(a)**

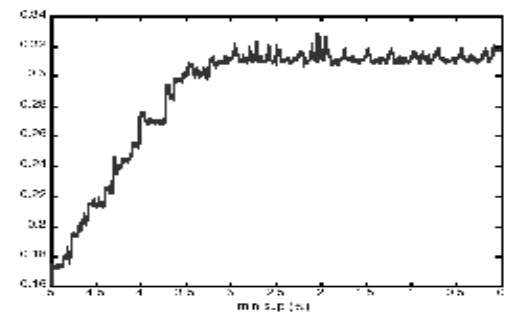

**(b)**

*Figure 3: Effect of decreasing minimum support value on a FP-tree. (a) FP-tree nodes increase when minimum support value decreases. (b) FP-tree construction time increases when minimum support value decreases. The experiment is performed on the synthetic data. 10K transactions are used in this experiment.*

({L10})→({ST}, {129.138}, {EV}) [40%, 67%], where 40% is the support, 67% is the confidence. The support value is calculated by dividing current node's occurrence count, 2 in this example, by total transaction count, 5 in this example. It indicates how important this rule is to profile the user behavior. We calculate the confidence by dividing current frequency node's occurrence count, 2 in this example, by the node's total occurrences, 3 in this example. It shows how strong this rule is. By following current node's link chain, we con get the second tree node {L10}:1. The corresponding association is {L10}->({ST},{ET}) [20%,33%], which is generated in the same way. This process continues until the null link chain pointer is met.

## 3.2 FP-TREE BASED PATTERN MATCHING ALGORITHM

Two techniques are utilized to build the rule monitor. To indicate the anomaly of a new transaction, we designed a novel FP-tree based pattern matching algorithm. And an alert accumulating algorithm is used to lower the false alarm and to detect a set of fraudulent transactions with low suspicious values.

For each frequent item $t_i$ in an FP-Tree, we should find all the prefix paths of $ti$, which is the sub patterns base under the condition of $t_i$'s existence. These prefix paths do not include all the possible patterns containing $t_i$. Since, some patterns are filtered by minimum support. An incoming transaction having more than one item in the header table means that this transaction matches the single frequent pattern in some level. To discover in what extent the new transaction matches the association rules, we need to walk through the FP-tree by following the header table link chains to compare that transaction to every sub pattern base.

```
double SimMatch(T) {
  sim = 0.0;
  for each item ti in T {
    if ( found headtablelink, in the head table ) {
      sim_credit = 0.0;
      for each tree node in N_i^j in headtablelink,
        if (P^j(ti) ⊆ T )
      sim_credit += G(N_i^j .s, N_i^j .c) x weight(ti);
      sim += sim_credit;
      }
    }
  return sim;
  }
```

The above pseudo code shows the pattern matching algorithm for calculating the similarity between a new transaction and the user's FP-Tree. Suppose $T = \{t1, t2, \ldots, tn\}$ is an incoming transaction. For each frequent item $t$ we calculate a similarity credits $sim\_credit(ti)$. by the following steps: Search $ti$ in header table. If not find, $sim\_credit(ti) = 0$. Otherwise, follow the link chain to the first tree node $Ni$ 1 containing $ti$. A conditional frequent pattern (set of frequent items) can be obtained by following $Ni$ 1's parent link until it reaches the root. Let's denote the attributes set in this pattern as $P1(ti)$. If $P1(ti)$ T, $sim\_credit(ti)$ is increased by a credit function, $G(s, c)$, based on the support and the confidence of $Ni$ 1. In our implementation, we used an entropy like function: $G(s, c) = -sx \log_2(1+\varepsilon-c)$, where $s$ is the support, $c$ is the confidence, $s \leq c \leq 1.0$, $\varepsilon$ is a real number used to specify the upper boundary of function $G$. The function is chosen by the intuition that if a new transaction matches with a rule having larger probability it is more likely to match the user's behavior, and the confidence value is emphasized.

Since different kinds of frequent items are of different importance to profile a user or a system. A weight function, weight($ti$), was used to give various stress to the different item types. So we increase $sim\_credit(ti)$ by $G(s, c)$ x weight($ti$) instead of $G(s, c)$. The weight function could be a fixed look up table, which maps the different item types to different weights. It is also possible to use a neural network to train real data to get the optimized look up table.

After $N_i^1$ is processed, we follow the link chain of $N_i^1$ to reach the second tree node, $N_i^2$, containing $t_i$. For that tree node we do the same thing as to $N_i^1$. The process is stopped when the $N_i^j$'s link chain pointer is null. The similarity value for $T$ is computed as:

$$sim(T) = \sum_{i=1}^{n} sim\_credit(t_i),$$

where $sim\_credit(t_i) = weight(t_i)$ x $\sum_j G(N_i^j.s, N_i^j.c)$, $j$ is the index of the node containing $t_i$ found by following the head table link of $t_i$.

The $sim(T)$ would be mapped into a corresponding suspicious value. Similarity value represents the extent that a new transaction is comparable to the customer behavior patterns. The minimum similarity value is zero, which means not a single rule is matched between the new transaction and the user pattern. It indicates the highest suspicious level. By contrast, the larger similarity value means a smaller suspicious level.

### 3.3 ALERT ACCUMULATING ALGORITHM

We can setup a set of thresholds to fire corresponding fraud warnings. Technically, it is possible to mis-classify some legal transactions, which do not follow the customer's normal behavior. Since the frequent items are filtered by min support before creating the FP-tree. Therefore the tree is not completed, the very unusual patterns are not collected in the tree. Moreover, a customer could also suddenly change his or her behavior. Another important issue is that in order to minimize the fraud detection cost, the purchased amount is a factor of firing alarm. For a very small purchase amount, for example .50$, even it is highly suspicious, fire an alarm is not economical. Since the objective of fraud detection action is to minimize the total cost. The cost model section explains this issue in detail. By using a suspicious threshold for a single transaction, a sequence of fraud transactions with low purchasing amounts could be missed.

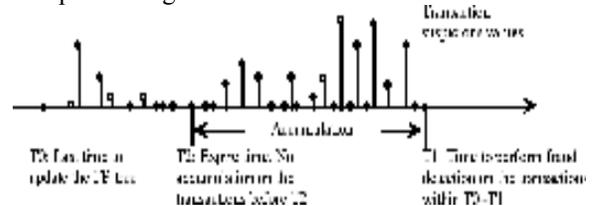

*Figure 4: Suspicious values within accumulation window. The height of a point indicates the*

*suspicious value of this transaction. T0 is the moment that the FP-tree is most recently updated. T1 is the current time. T1 and T2 is the boundary of the accumulation window.*

To solve these problems, we use a novel technique to accumulate the warnings from a set of new transactions.

We calculate the alert values for a set of transactions by accumulating their suspicious values. The transactions to be processed could include all the new transactions after last FP-tree updating, or we can use an expiring time window to specify the transactions we would like to accumulate. Then by comparing the alert sum instead of the single alert against a set of thresholds, a corresponding fraud alarm would be fired. The threshold set is decided by the detection sensitivity specified by the user.

We accumulate all transactions within the specified time window (T2~T1), as shown in Figure 4, by adding them together with the same emphasis. A simple step function is the most straightforward expiring function. Some examples of the expiring functions are shown in Figure 5. For a step function, if time of a transaction, $t$, is larger than T2, $f(t)$ equals to 1, else $f(t)$ equals to 0. We also can emphasize the transactions that happen more recently by using either the nature logarithm function or the polynomial functions.

The fraud alert value is calculated by: $AlertValue = \sum_{i=1}^{n} (s(T_i) \times f(T_i) \times amount(T_i))$, where $T_i$ is a transaction in the accumulation window, $s(T_i)$ is the suspicious value of $T_i$, $f(T_i)$ is the expiring weight of $T_i$ specified by expiring function, $amount(T_i)$ is the consumed money of transaction $T_i$. It is reasonable that several highly suspicious transactions having very low consumed points will not fire an alert, by contrast, only one highly suspicious transaction having very high consumed points could fire an alert.

## 4. PERFORMANCE STUDY
### 4.1 DATA PREPARATION

The real commercial database for performance evaluation is far more than possible. The only available real time data for our experiments is a set of American Express Credit Card transactions of an individual user within three and a half years. It is inadequate to measure the system performance since the real time behaviors could be various from user to user. To evaluate the system's effectiveness on different kinds of user behaviors, we created a tool to simulate different kinds of user behaviors. We then tested and evaluated NMT-FDS and other well known algorithms (Naive Bayes, C4.5, BP, SVM) on the same data sets.

Diversity real time user behavior requires a powerful and flexible simulator. To generate the various user patterns, profile driven is an effective solution, since the profile could be creatively designed in order to represent complex patterns. Our simulator is similar to the one proposed [8] by Chung, whose functions are parsing the user profile and modeling the user behavior by following the rules defined in the profile file.

The simulated data has the very similar formation as the American Express data. Table 1 is an example of the simplified record coming from the simulated data sets. Two different types of user profiles are used in our experiments to simulate the relatively regular behaviors and the relatively irregular behaviors. An example of regular behavior profile used in experiments is: most of the transactions come from a small IP group, most transactions take place at weekend, transactions are purchasing a group of products, the transactions likely occur in the evening. An example of irregular behavior is: transactions come from dynamic IPs, most transactions take place at weekend and transactions are purchasing a group of products.

For both of the profiles, we create 3000 legal transactions and 50 fraudulent transactions as a labeled training set to train the supervised classification algorithms. Our approach takes all 3050 transactions as an unlabeled data set, that is to say, we don't know the fraudulent and legitimate when we create the user profile. The trick is that the fraudulent behavior with very low occurrence ratio will be filtered out when we create the FP-tree. We also create 3000 legal transactions and 20 fraudulent transactions as testing data.

### 4.2 EXPERIMENT RESULTS

Figure 5 on the following page shows the performance comparison among BP, NB, SVM and NMTFDS on both real time and simulated data. We use the ROC (Receiver Operating Characteristics) curve to evaluate these classification algorithms.

Figures 5(1)–5(3) show the ROC curves of three algorithms on relatively regular data. The areas under the ROC are very close to 1, which means they work well to differentiate fraudulent data against regular behavior data. Figures 5(4)–5(6) show the ROC curves on a test data set of the irregular behaviors. Figure 5(7)–5(9) show the ROC curves on a data set of an American Express user. Experimental results show that our algorithms, NMT-FDS works much better on irregular data than other algorithms. The reason is that dynamic IPs in irregular behavior data set introduces noises which largely hamper accurate learning of NB, SVM and BP. Since noises or rarely attributes. Such as dynamic IPs would be automatically filtered by the min-support, FSD is capable to work on the irregular data. A tree based classification algorithm, C4.5, has also been studied. It has a very similar performance to NMT-FDS. However the supervised learning makes C4.5 inadequate for adaptively profile updating.

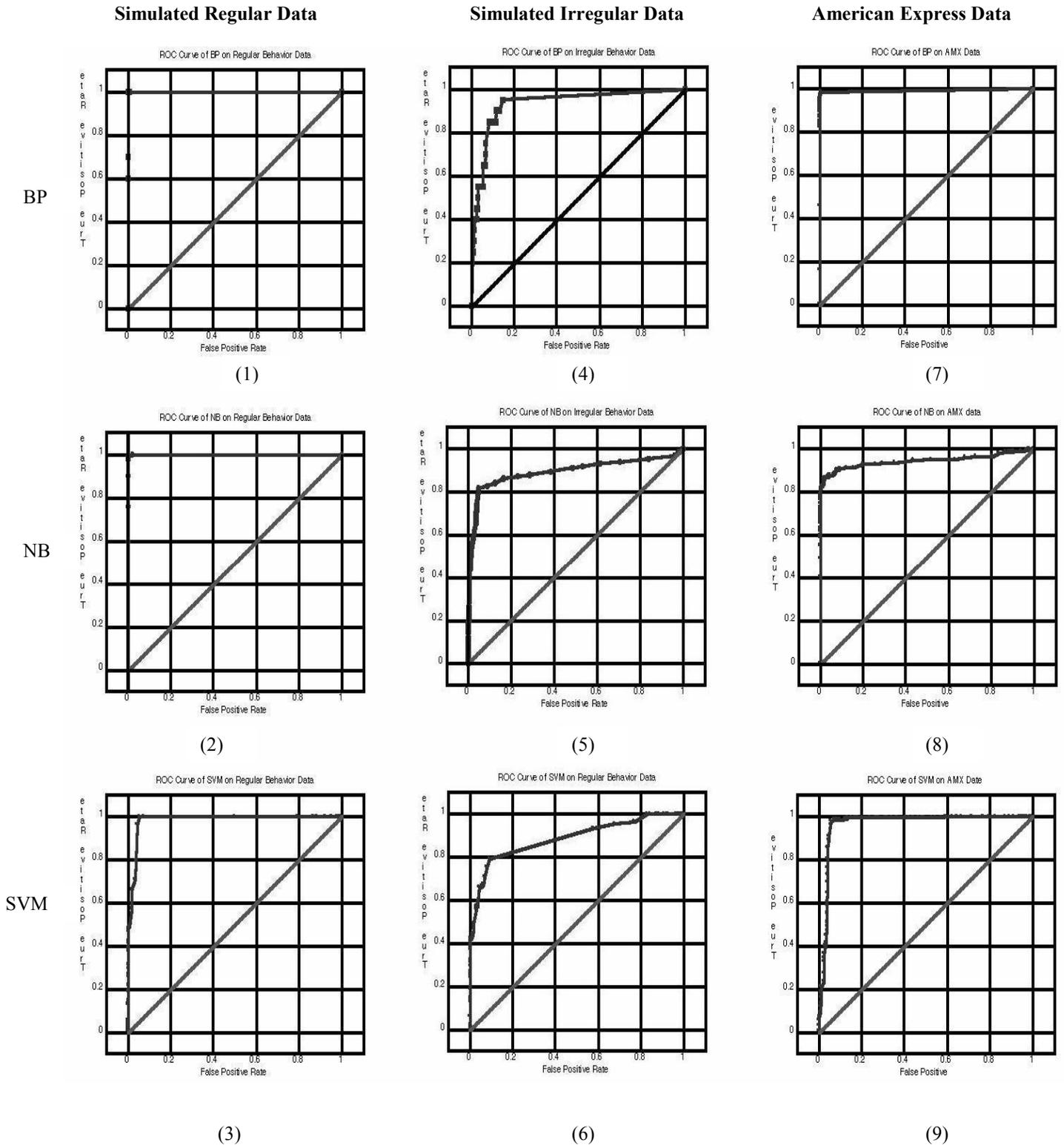

Figure 5: ROC curve comparison among the different algorithms. (1)–(3) show the ROC curves of different three algorithms on relatively regular data. (4)–(6) show the ROC curves on relatively irregular data. (7)–(9) show the ROC curves on a data set of a single American Express user.

ROC curve is a very convenient way of comparing the performance of classifiers, however it cannot show the misclassification cost. To measure the benefit of detecting fraud, we used a Cost Model introduced by Chan  a more realistic model [5], to accompany the different classification outcomes. A cost model formulates the total expected cost of fraud detection. It considers the trade off among all relevant cost factors and provides the basis for making appropriate cost sensitive detections. The detection outcome is one of the following: *hit, false alarms, miss*, and *normal*. They are outlined in Table 2.

| Prediction | Fraud | Legal |
|---|---|---|
| Alert | *Hit* | *False Alarm* |
| No alert | *Miss* | *Normal* |

*Table 2: Outcome Matrix for fraud detection*

## 5. CONCLUSIONS AND FUTURE WORK

We have presented a framework for detecting fraudulent transactions in an online system. We describe the major modules of the framework and the related algorithms in detail. A prototype of the fraud detection system has been built to evaluate our algorithm. A profile driven simulator is designed to generate transaction data representing various behavior patterns in order to evaluate the performance of our algorithm. Comparisons are performed among NMT-FDS, C4.5, NB, BP and SVM. Table 3 shows a summary of the qualitative comparison. Our system generates fraud alarm accurately and efficiently on both the simulated and real data. Unsupervised training and self adjustment to changing user behavior make our system effective for monitoring online transaction systems and provide fraud detection and protection.

In the future, we will extend our system to detect system level fraud by utilizing an approximate weighted tree matching algorithm. Inter-transaction behavior mining is also planned to enhance the performance of individual user profiling. Our future work also includes optimizing pattern matching algorithm, optimizing weight selection for different types of rules by real data training and expanding our algorithm to real-time fraud prevention.

| Algorithms | Effectiveness | Scalability | Speed | Training Data | Adaptability |
|---|---|---|---|---|---|
| NB | Various | Good | Good | Labeled | Poor |
| C 4.5 | Good | Poor | Fine | Labeled | Poor |
| BP | Fine | Good | Poor | Labeled | Poor |
| SVM | Fine | Good | Poor | Labeled | Poor |
| NMT-FDS | Good | Good | Fine | Unlabeled | Good |

*Table 3: Summary of qualitative comparison of algorithms.*

Table 3 shows *Effectiveness* highlights the overall predictive accuracy and performance of each algorithm. *Scalability* refers to the capability to construct a model effectively given large data sets. *Speed* indicates the efficiency in model construction. *Training data* shows the training data in model construction. *Adaptability* refers the capability and efficiency to adjust the model to follow the changes of the user behaviors.

## Author's Biography

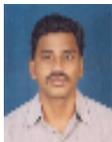**P. Srinivasulu** received his B.Tech from Acharya Nagarjuna university, Guntur, AP in 1994 and completed post graduation from Jawaharlal Nehru Technological University, Hyderabad in 1998.  He is currently pursuing Ph.D from Acharya Nagarjuna University, Guntur and working as Assistant Professor in V R Siddhartha Engineering College, in the Department of Computer Science and Engineering, Vijayawada, Andhra Pradesh. His research interest includes Data Mining and Data Warehousing, Computer Networks, Network security and Parallel Computing. He has more than thirteen years of experience in teaching in many subjects, industry and in research. He is the member of Indian Society of Technical Education (ISTE) and also member of Computer Society of India(CSI). He has many publications in National and International conferences.  He was selected for the Journal of Who is who.

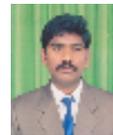**Mr. J Ranga Rao r**eceived his B. Tech in Computer Science and Engineering, J N T University, Hyderabad in 2005. He also received his M. Tech in Computer Science and Engineering from J N T University, Kakinada in 2008. Presently working as a Lecturer in the department of computer science and engineering of V R Siddhartha Engineering college, Vijayawada. He is an associative member of IST, India.

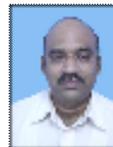**Dr. I Ramesh Babu** received his Ph. D in Computer Science from Acharya Nagrjuna University, Guntur, M. E in Computer Engineering from Andhra University, B.E in Electronics and Communication Engineering from University of Mysore. He is currently working as Professor in the department of Computer Science, Nagarjuna University. Also he is Senate member of the same university from 2006. He held many positions in Acharya Nagarjuna University as Head, Director-Computer Center, Chairman-Board of studies. He was a special officer, convener of ICET, Andhra Pradesh. He is also member of board of studies for the other universities. His areas of interest include image processing, computer graphics, cryptography, artificial intelligence, and network security. He is a member of IEEE, CSI, ISTE, IETE, IGISS, and Amateur Ham Radio (VU2 IJZ). He is currently supervising ten Ph. D students who are working in different areas of image processing and artificial intelligence. He has published 35 papers in international journals and conferences.